\documentclass[fleqn,usenatbib]{mnras}

\usepackage{mathptmx}
\usepackage{txfonts}

\usepackage[T1]{fontenc}

\DeclareRobustCommand{\VAN}[3]{#2}
\let\VANthebibliography\thebibliography
\def\thebibliography{\DeclareRobustCommand{\VAN}[3]{##3}\VANthebibliography}

\title[NS LMXB recurrence times]{On the recurrence times of neutron star X-ray binary transients and the nature of the Galactic Centre quiescent X-ray binaries}

\author[Maccarone et al.]{
Thomas J.~Maccarone,$^{1}$\thanks{E-mail: thomas.maccarone@ttu.edu (TJM)} 
Nathalie Degenaar,$^{2}$, Bailey E.~Tetarenko,$^{1}$, Craig O.~Heinke$^3$, Rudy Wijnands$^2$,
\newauthor Gregory R.~Sivakoff $^3$
\\
$^{1}$Department of Physics \& Astronomy, Texas Tech University, Lubbock TX, 79409-1051, USA\\
$^{2}$ Anton Pannekoek Institute for Astronomy, University of Amsterdam, Amsterdam, The Netherlands\\
$^{3}$ Department of Physics \& Astronomy, University of Alberta, Edmonton, Alberta, Canada
}

\date{Accepted XXX. Received YYY; in original form ZZZ}

\pubyear{2021}

\begin{document}
\label{firstpage}
\pagerange{\pageref{firstpage}--\pageref{lastpage}}
\maketitle

\begin{abstract}
The presence of some X-ray sources in the Galactic Centre region which show variability, but do not show outbursts in over a decade of monitoring has been used to argue for the presence of a large population of stellar mass black holes in this region.  A core element of the arguments that these objects are accreting black holes is the claim that neutron stars (NSs) in low mass X-ray binaries (LMXBs) do not have long transient recurrence times.  We demonstrate in this paper that about half of the known transient LMXBs with clear signatures for NS primaries have recurrence times in excess of a decade for outbursts { at the sensitivity of MAXI}.  We furthermore show that, in order to reconcile the expected total population of NS LMXBs with the observed one and with the millisecond radio pulsar (MSRP) population of the Galaxy, systems with recurrence times well in excess of a century { for outbursts detectable by instruments like MAXI} must be the dominant population of NS LMXBs, and that few of these systems have yet been discovered.
\end{abstract}

\begin{keywords}
accretion,accretion discs -- binaries:close -- stars:neutron -- X-rays:binaries
\end{keywords}



\section{Introduction}

Low mass X-ray binaries (LMXBs) are binary systems with low mass donor stars (i.e. $\lesssim 1 M_\odot$) and black hole or neutron star (NS) accretors.  Most of these systems are susceptible to ionization instability effects that make their accretion disks have transient outbursts, which are quite similar to the dwarf nova instabilities seen in cataclysmic variable stars \citep{1989ApJ...343..241M, Lasota2001}, with modifications to the temperature profiles of the disks being required due to the stronger irradiation of the disks by X-rays from their inner regions in the X-ray binaries \citep{vp1996,Dubus2001}, resulting from to the deeper gravitational potential wells of their accretors.  { Outbursts may also, in principle, be triggered by variations in the mass transfer rate from the donor star \citep{Bath}, and these effects may be responsible for factors of few variability in quiescence in accretion disks around black holes and neutron stars \citep{Cantrell,Cackett,Bernardini} and almost certainly are responsible for some long, weak outbursts in accreting white dwarfs \citep{Rivera2021}.}

Among the known samples (which are strongly biased toward sources bright enough to trigger all-sky instruments), typical black hole X-ray binaries show peak luminosities in their transient outbursts of $10^{37-39}$ erg/s \citep{WATCHDOG}, while typical NS LMXBs show peak luminosities about a factor of 10 fainter \citep{YanYu}.  The peak luminosities show a clear correlation (with some scatter) with the orbital period of the binary \citep{Wu2010}, another phenomenon analogous to that in the cataclysmic variables \citep{Warner1987}. This is expected if a critical mass density must be reached in the outer accretion disk to trigger an outburst, since the larger disks will then have a greater total amount of mass transported.  The quiescent luminosities of these systems also correlate reasonably well (again, with some scatter) with the orbital periods \citep{Garcia2001, Reynolds2014}, and are typically 3-8 orders of magnitude lower than those in near the peaks of the outbursts.  This is expected if the accretion rate in the inner disk scales with the mass transfer rate into the outer disk for all quiescent X-ray binaries, and the mass transfer rate correlates with the orbital period, as expected from binary evolution scenarios (see e.g. \citealt{KKB}).

A variety of classes of surveys have the potential for finding X-ray binaries in quiescence -- X-ray surveys \citep{GBS},  optical emission line surveys \citep{Casares2018},  flat spectrum radio source surveys \citep{tjm2005,ngvla}, binary wobble \citep{Gould2002,Breivik2017,Thompson2019,Jayasinghe2021} all hold potential for finding these systems.  At the present time, discovery of X-ray binaries in quiescence has been mostly limited to sources in or near globular clusters (e.g. \citealt{Rutledge02a,2003ApJ...598..501H,tetarenko2016,Bahramian20}), while X-ray all-sky monitors, which have been deployed for most of the past 50 years, and continuously since the launch of RXTE in late 1995, can detect bright (i.e. $L_X>10^{37}$ erg/sec ) outbursting X-ray binaries over most of the Milky Way. 

A substantial number of low-luminosity outbursting and quiescent X-ray sources have been identified near the Galactic Centre \citep{Muno03,Muno05,Degenaar09,Degenaar12}, which is one of the regions with sufficiently deep X-ray data in which to detect quiescent X-ray binaries and with sufficient monitoring by pointed instruments to detect faint ($\sim 10^{35}$ erg/sec) outbursts.   
Sixteen of the quiescent objects in this data set 
have been argued to be  
quiescent stellar mass black holes \citep{Hailey2018,Mori} and it is further argued that these represent a set of gas-capture binaries, with $\sim 300$ such black hole binaries in the nuclear star cluster \citep{Tagawa2020}.  { These objects have X-ray spectra inconsistent with magnetic cataclysmic variable stars, which show strong iron emission lines \citep{Hailey2018}.  They are brighter than most, but not all, non-magnetic CVs, and while the non-magnetic CVs have weaker emission lines than the magnetic ones, even the non-magnetic CVs would have stronger iron emission than these systems \citep{Mori}.}  None has shown an X-ray outburst (although most do show substantial non-outburst variability).  A core argument 
of \citet{Hailey2018}
is that the lack of bright outbursts from these systems demonstrates that they are extremely likely to have black hole accretors.  { Specifically, \citet{Hailey2018} argue that they rule out neutron-star, low-mass X-ray binaries as candidates for these objects because the large outbursts of accreting neutron stars have recurrence times of about 5-10 years. That statement is the only argument made by \citet{Hailey2018} that the systems they identify as black hole accretors cannot be neutron stars.  The same arguments are repeated in \cite{Mori}, with the latter paper acknowledging that not all neutron stars which have large outbursts recur on timescales less than 10 years, but treating the more infrequently recurrent systems as irrelevant outliers.}

In this Letter, we show that multiple lines of evidence indicate that the lack of outbursts over 
2-3 decades 
cannot be used as reliable evidence that these sources are black holes.  
We show 
by reference to the literature 
that (1) a large fraction of known NS LMXBs shows outbursts with long recurrence times, and that (2) the sample of known NS LMXBs is much smaller than predicted by binary evolution models and much smaller than the size of the population needed to produce the observed population of MSRPs. { In neither of the papers claiming a population of quiescent black hole X-ray binaries in the Galactic Center region \citep{Hailey2018,Mori} do they define a ``large'' outburst quantitatively, but from context, it is clear that they are referring to the kinds of outbursts that can trigger all-sky instruments.  For the purposes of this paper, we take that to be a luminosity of $2\times10^{36}$ erg/sec or more, which would trigger typical all-sky monitors at Galactic Center distances. We define long recurrence times to be recurrence times of at least 10 years, again based on the claim by \citealt{Hailey2018} that recurrence times longer than 10 years are indicative of black holes, rather than neutron star accretors.  Within the context of the above definitions, we show that neutron star accretors frequently have long recurrence times for their large outbursts.}

\section{Transient properties of outbursting systems}

When looking at the sample of transients seen with the RXTE All-Sky Monitor (ASM) above 100 mCrab\footnote{This corresponds to 1.5$\times10^{37}$ erg/sec at 8 kpc.}, \citet{YanYu} find that 14 different NS accretor objects had a total of 68 transient outbursts.  About 72\% of the outbursts came from three sources, Aql~X-1, 4U~1608-52 and MXB~1730-33 (the Rapid Burster).  The former two objects can both be inferred to have subgiant donor stars based on their orbital periods of 18.9 and 12.9 hours, respectively.  This leads to mass transfer driven by the expansion of the donor star, which gives a higher mass transfer rate than is possible in most shorter period systems \citep{KKB}; at the same time, the accretion disc radii are not dramatically bigger than for main sequence donor systems, so the ratio of the accretion rate to the critical accretion rate for stability \citep{Lasota2001} is higher, and the duty cycle of the outbursts is higher as well.  Due to the Rapid Burster's location in Liller 1, a highly reddened globular cluster, its donor star cannot be studied.  

It is thus clear that the number of {\it outbursts}  seen from NS LMXBs is dominated by the systems with a short recurrence time, even if the number of known {\it outbursting objects} is not.   It is likely because of these systems that a point of astronomical lore has developed that the NS LMXBs are predominantly systems with short recurrence times.  It is also true that there are many LMXBs with NS accretors that appear to be persistent accretors.\footnote{It is likely that most are persistent, but a small fraction of these objects may be systems with outburst durations that exceed duration of the history of X-ray astronomy.}  The frequently recurrent transients and the persistent sources generally dominate the total number of publications in the literature, because the amount of useful data for these sources is so much greater than for the long recurrence time transients.  

It is clearly {\it not} the case that the number of {\it systems} is dominated by short recurrence time transients.  Even in the \citet{YanYu} study, which considered data only from RXTE, 5 of the 14 studied systems show only a single outburst, and 4 more objects show only two outbursts.  Therefore, given the 16-year long mission lifetime for RXTE, the majority of the NS X-ray binary transients that reached 100 mCrab in the RXTE era were systems with recurrence times of at least 5 years, and more than  1/3 of the systems have recurrence times of at least 8 years. {\it By number of sources}, rather than by number of outbursts, the relatively long recurrence time transients are quite common.  When attempting to interpret quiescent systems, it is the number of sources that must be considered, rather than the number of outbursts.  In fact, if there is a substantial number of single outburst transients in the sample, then they should be assumed to represent a much larger underlying population of transients that have recurrence times longer than the duration of existing surveys.  Similarly, the classical novae are generally believed to have typical recurrence times of thousands to tens of thousands of years based in part on models and in part on the fact that so few of them recur, meaning that the typical recurrence timescale must be significantly longer than the roughly one century duration over which optical monitoring programs have been able to identify them.  Also similarly, there is a small subset of ``recurrent novae'' which can recur as quickly as yearly \citep{Bode2011}.

We can further examine the known transient population to illustrate the need for a large population of long recurrence time transients.  We start from the \citet{Liu_catalog} catalog of X-ray binaries, and add to it, in Table 1, a set of NS LMXBs that showed their first transient outbursts after that catalog was finalized.  We limit the \citet{Liu_catalog} sources to those which are identified as transients, and which show evidence for either bursts, pulsations, or quiescent soft thermal emission that can be used to classify them clearly as NS LMXBs.  We furthermore exclude sources which are in sufficiently crowded regions that either outbursts from them can be missed (e.g. 1M~0836-425, which is very close to a bright accreting pulsar) or where  the poor positional accuracy of some early all-sky missions means that there may be several sources with individual outbursts or a recurrent transient (e.g. sources in the Galactic Center region that do not have sub-arcminute follow-up). We also exclude sources for which the peak fluxes were less than 30 mCrab, since those could not be discovered with the RXTE ASM.  Within the excluded sample are likely to be several additional NS X-ray binaries with long recurrence times, but our goal in this Letter is simply to illustrate that a large population of long recurrence time transients exists.  Where necessary to establish the long recurrence times for a source, we supplement literature data (mostly from \citealt{YanYu,Lin18}) with data from the Monitor of All-sky X-ray Image (MAXI -- \citealt{2009PASJ...61..999M}).  

Then, using these literature compilations, and MAXI data, we assess the recurrence times of the transient systems.  { The MAXI data reach a detection limit of 1 mCrab in one week \citep{Isobe}.  Most of the other all-sky instruments are somewhat less sensitive than MAXI, but if the source was detected before MAXI's launch, then MAXI extends the baseline over which its outbursts have been studied to at least 10 years, so the poorer sensitivity of some of the earlier monitors is not a concern for the purposes of establishing that recurrence times for bright outbursts are long.  }

In Table \ref{longrecurtable}, we present a list of  21 long recurrence time transients that have survived our cuts.  The \citet{Liu_catalog} catalog from which we started had 45 LMXBs that showed clear evidence for NS primaries, and then we have added 6 more sources discovered after the production of that catalog.  From this analysis, we can say that {\it at bare minimum} 40\% of the transient LMXBs have long ($>$10 year) recurrence times.  The number {\it must} be much larger due to the need for ultralong ($>50$ year) recurrence time systems which have not yet been discovered to match the population synthesis numbers and long period MSRP numbers, and also to explain the presence of so many systems with only a single known outburst, as well as the continued discoveries of new NS LMXBs making their first outbursts.

There is considerable and clear evidence that some X-ray binaries and cataclysmic variables show clustering in the times of their outbursts.   Thus, beyond the clear evidence for long recurrence times among known transients, there is good reason to believe that some of the known transients with short recurrence timescale may alternate between epochs of longer and shorter recurrence times, with selection biases obviously favoring them being in the short recurrence time phases of their lifetimes when we detect them.  Among NS X-ray binaries, Cen~X-4, which showed extremely bright outbursts in 1969 and 1979 \citep{CSL}, but has been quiescent since 1979, provides a good example.  
Among the black hole X-ray binaries, there are several objects which were undiscovered until the 1990s, and have shown several outbursts since then that were sufficiently bright and sufficiently long duration that they would have been seen with earlier monitors (e.g. XTE~J1550-564 and GRO~J1655-50 -- \citealt{BlackCAT,WATCHDOG}).  In cataclysmic variables, the clustering of outbursts has been seen in a few cases \citep{Kato2002,Wenzel,1990AJ.....99.1941B} as well, with some work arguing that magnetic activity cycles in the donor stars are the likely cause \citep{1990AJ.....99.1941B}. Clustered outbursts can lead to objects alternating between epochs of shorter and longer recurrence times.  Given observational biases toward systems which do outburst, it is likely, then, that this leads to an underestimate in the typical recurrence times based on empirical data.  { The observed clustering of the outbursts is likely to be due to short timescale effects in either the donor star or the accretion disc;  on longer timescales, of tens to hundreds of Myr, the outburst recurrence times may also change due to more fundamental, secular changes in the mass transfer rates.}

\begin{table*}
\begin{tabular}{|l|l|l|l|l|}
\hline
  \multicolumn{1}{|c|}{Name} &
  \multicolumn{1}{c|}{RA} &
  \multicolumn{1}{c|}{Dec} &
  \multicolumn{1}{c|}{Comments} &
  \multicolumn{1}{c|}{References}
 \\
\hline
  EXO 0748-676 & 07 48 33.300 & -67 45 00.00 & single long outburst & (1)  \\
  XTE J0929-314 & 09 29 20.190 & -31 23 03.20 & Single outburst in 2002 & (2,3) \\
  4U 1456-32 & 14 58 22.000 & -31 40 08.00 & Cen X-4,  outbursts 1969 \& 1979, none since & (4) \\
  MXB 1659-298 & 17 02 06.500 & -29 56 44.10 & 3 outbursts in \citet{Lin18}, $\sim 1$ per 14 years& (1)  \\
  XTE J1701-462 & 17 00 58.450 & -46 11 08.60 & single outburst, very bright and long duration& (5) \\
  XTE J1723-376 & 17 23 38.700 & -37 39 42.00 & single outburst with RXTE, not seen with MAXI& (6)  \\
  4U 1730-220 & 17 33 57.000 & -22 02 07.00 & outbursts in 1972 and 2021 &  (7,8) \\
  KS 1731-260 & 17 34 13.470 & -26 05 18.80 & single outburst from 1989-2001& (9,10)  \\
  EXO 1747-214 & 17 50 24.520 & -21 25 19.90 & single outburst & (11)  \\
  2S 1803-245 & 18 06 50.720 & -24 35 28.60 & outbursts in 1976 and 1998& (12) \\
  XTE J1807-294 & 18 06 59.800 & -29 24 30.00 & two outbursts during RXTE mission, so $\lesssim 0.1$/yr&(1) \\
  GS 1826-238 & 18 29 28.200 & -23 47 49.12 & single very long outburst with reflaring&  (13)\\
  HETE J1900.1-2455 & 19 00 08.650 & -24 55 13.70 & single outburst &(1)\\
  4U 1905+000 & 19 08 26.970 & +00 10 07.70 & Off since at least 1990\\
  &&&Not clear if one long outburst or \\
  &&&some clustering of outbursts& (14)  \\
  XTE J2123-058 & 21 23 14.540 & -05 47 53.20 & single outburst& (1)  \\
  \hline

  MAXI J1647-227& 16 48 12.32& -23 00 55.2   & single outburst & (15)  \\
  MAXI J1421-613 &14 21 37.2 & -61 36 25.4  &  possible outburst in 1970s, outburst in 2014 &(16,17) \\
  Swift J185003.2-005627 & 18 50 03.2 &-00 56 27 & single outburst & (19)   \\
  MAXI J1621-501 &16 20 22.01& -50 01 11.6  &single outburst& (20,21)  \\
  Swift J181723.1-164300 & 18 17 23.1 & -16 43 00 & single outburst &  (22)\\
\hline\end{tabular}
\caption{LMXBs with NS accretors that have shown outbursts separated by at least 10 years in time, or single outbursts bright enough that they would have triggered all-sky monitors over the period in which all-sky monitors existed (i.e. since 1996). Objects above the horizontal line are in the \citet{Liu_catalog} catalog, while objects below the line were discovered since its publication.  All the objects below the line show Type~I X-ray bursts \citep{2020ApJS..249...32G}.  References: (1) \citet{Lin18} (2) \citet{Galloway2002} (3) \citet{Cartwright} (4) \citet{CSL} (5) \citet{Homan1701}(6)\citet{YanYu} (7) \citet{Cominskey1978} (8) \citet{2021ATel14757....1I}  
    (9) \citet{1989IAUC.4839....1S} (10) \citet{2001ApJ...560L.159W} (11) \citet{1985IAUC.4058....2P} (12) \citet{Cornelisse2007} (13) \citet{Ji2018}  (14) \citet{Jonker2006} (15) \citet{MAXI1647} (16)\citet{MAXI1421} (17) \citet{2015PASJ...67...30S} (18) \citet{Falanga2006} (19) \citet{2011GCN.12083....1B} (20) \citet{2017ATel10869....1H}  (21) \citet{2017ATel10874....1B} (22) \citet{2017GCN.21369....1B} }
    \label{longrecurtable}
\end{table*}

An essential point here is that what matters for using long recurrence times to infer the nature of a compact object is what fraction of the {\it underlying population} of sources have long recurrence times.  It does not matter whether the fraction of the transient NS LMXBs observed in a single snapshot, or even over the $\sim 60$ year history of X-ray astronomy is dominated by a particular class of objects.  The sources observed in single snapshots are very strongly biased toward high duty cycle transients.

Given the presence of a substantial population of NS LMXBs with only a single observed outburst, the underlying population's size cannot be directly assessed from the data, without some modelling assumptions.  Still a bit more than half of known NS transients have shown only a single outburst, and the vast majority of the expected population remains quiescent and undiscovered unless the total NS LMXB population is a few hundred objects.

\section{What the overall Galactic Population of NS LMXBs tells us}

Not only does the population of known NS LMXBs includes a substantial contribution from systems with long recurrence times, it is furthermore likely that this population is the dominant population of NS LMXBs.   Numerous population synthesis works find that 2000--10000 NS LMXBs should exist in the Milky Way \citep{Kalogera1998,Kiel2006,vanHaaftenLMXB2015}.  The range of estimates from first principles binary population synthesis depends largely on treatment of natal kicks of the NS and on the common envelope phase of binary evolution, both of which are relatively poorly constrained from the existing data, and 
are extremely challenging problems to approach using theoretical methods.  If, on the other hand, it were assumed that there were no long recurrence time NS X-ray binaries (and hence that the known sample of about 80 objects represents the whole Milky Way sample), then the total number of NS X-ray binaries would be underpredicted by a factor of $\sim50$.

An approach which is more robust to model assumptions, but also rougher, is the  comparison of the populations of MSRPs with the populations of NS LMXBs.  The MSRPs must be spun up in LMXB evolutionary phases.  From this approach, if the population level of the LMXBs is in steady state (which is likely, given that the Milky Way's star formation rate is roughly constant over the past 8 Gyr -- \citealt{Snaith2015}). The total MSRP population is about 40,000 \citep{LorimerLRR}, and these systems are expected to live for a Hubble time, so that the typical age is about $5\times10^9$ years.  Then, the number of NS LMXBs, $N_{\mathrm{NS-LMXB}}$, can be expected to be 40000 $(\tau_{\mathrm{LMXB}}/5\times10^9 {\rm yr})$.
Then, to have $\sim$200 LMXBs, so that only a small fraction have been discovered, requires that the LMXBs have characteristic lifetimes of $2\times10^7$ years over which they show phases of bright X-ray emission.  Such short lifetimes are at odds with binary evolution calculations that show that LMXB lifetimes as sources which are at least sporadically bright are typically at least $\sim10^8$ years \citep{Podsi2002}.

For relatively short period NS X-ray binary transients, \citet{Lin18} showed, using the orbital period-peak luminosity relations from \citet{Wu2010}, that all-sky surveys with 50 mCrab sensitivity should be complete within about 3 kpc.  The very faint X-ray transients (VFXTs) do have peak luminosities about 10 times fainter than this value, but MAXI's sensitivity is also about 1 mCrab, so MAXI should be complete to VFXTs within 3 kpc. { Thus, within 3 kpc, we have completeness to a luminosity limit well below that of ``large outbursts''.}.  Next, we determine by integrating the local stellar density, that about 5\% of the stellar mass of the Milky Way disk is within 3~kpc, if one assumes a Galactic scale length of 2.1~kpc, meaning that there should be 100-500 NS LMXBs within 3 kpc if the population synthesis estimates are correct. 

Only $\sim5$ of the sources with known distance estimates are found at distances closer than 3~kpc \citep{Jonker2004,Arnason2021}.\footnote{The number is approximate because there are many sources with distance uncertainties that allow them to be closer to or further than 3~kpc.}  Therefore, to be compatible with the theory derived both from best estimates and from expected ratios of MSRPs to LMXBs, there should be $\sim20-100$ undiscovered NS LMXBs for every one that is known.   The most likely manner in which to explain the lack of known nearby NS LMXBs is if there is a large population of transients which have yet to be discovered.  { Nearby undiscovered transients which are not VFXTs must have recurrence times longer than the 25 years over which continuous all-sky monitoring has been available, and nearby VFXTs must have recurrence times longer than the 12 years over which MAXI has been operating}.

Additionally, binary MSRPs predominantly have orbital periods of at least 1 day \citep{LorimerLRR}, and often have much longer orbital periods.  These long period binary pulsars are explained as arising from the X-ray binaries that evolve from shorter to longer periods via the expansion of their donor stars (e.g. \citealt{Podsi2002}).  These systems are likely to appear either as persistent X-ray binaries (like Sco~X-1 or Cyg~X-2) or as transients with long durations, long outbursts and very long recurrence times (like KS~1731-260 and XTE~J1701-462, both of which have shown outbursts of several years, and either of which has yet recurred)\footnote{The system MXB~1629-298 has similar outburst properties, but with an orbital period of 7.1 hours, it is an unlikely progenitor of a millisecond pulsar with a long orbital period, as it should be evolving to shorter period.}.  These objects thus likely exist in fairly large numbers, as transients with $\sim$ decade outburst durations and $\sim$ millennium or longer recurrence timescales (e.g. \citealt{Piro2002}); even if they recurred as frequently as once a century, then we would already have found $\sim$ half of them, given the $\sim$ 60 year history of X-ray astronomy.

\section{Quiescent NS LMXBs in Globular Clusters}
At low foreground extinction, NS LMXBs in quiescence show a signature of  thermal  cooling.  Their X-ray spectra are well fitted by a NS atmosphere model with a characteristic radius of about 10 km \citep{Rutledge}.  Many of these thermal emitters \footnote{These systems also often show power law tails to their spectra.} have been found in globular clusters \citep{2003ApJ...598..501H}, where the extinction is low, and dynamical interactions lead to the formation of new close binaries, a process which might  also be relevant in the innermost part of the Galactic Bulge \citep{2007MNRAS.380.1685V}.  The cluster 47~Tucanae, in particular, is both nearby, and sufficiently far from the ecliptic plane that it is continuously observable for most all sky monitors.  It has two quiescent LMXBs with clear spectroscopic evidence for being NS, but has never shown an X-ray outburst \citep{2003ApJ...598..501H,Bahramian2014}, and the cluster has also even been the subject of past Swift monitoring to search for VFXTs, which also have not yet been found.\footnote{See http://research.iac.es/proyecto/SwiftGloClu} M62 has 5 of these objects, and has never shown an X-ray outburst \citep{2003ApJ...598..501H,Bahramian2014}, and given the large number of candidates from which to have outbursts, even if a few outbursts were lost due to taking place while the cluster was close to the Sun, that would leave some objects which had not shown outbursts.  In the region near the Galactic Center, these soft components are typically not detectable with Chandra, because of the very high foreground absorption (i.e. $\sim 4\times 10^{22}$ cm$^{-2}$). 

The numbers of these quiescent NS LMXBs scale with stellar encounter rate \citep{Pooley03,2003ApJ...598..501H}, so we can use the stellar encounter rate calculations of \citet{Bahramian13} to estimate the number of quiescent NS LMXBs in the full system of Milky Way globular clusters, based on published observations of globular clusters. Twenty-four clusters with published {\it Chandra} observations are deep enough to identify most quiescent NS LMXBs (reaching $10^{32}$ erg/s).  These clusters contain 50 quiescent NS LMXBs and about 38\% of the dynamical interactions in Milky Way globular  clusters occur in this set of objects, assuming that the interaction rate follows the formulation of \citet{GammaVerbunt}.

Thus, we can estimate that there should be roughly 130 quiescent NS LMXBs within the Galactic globular cluster system.  (Note that not all quiescent NS LMXBs would be detected by this method; \citealt{Heinke05b} estimate that at least half of quiescent NS LMXBs would likely be missed by this method, as half show relatively hard quiescent spectra, and/or are significantly fainter).  On the other hand, all-sky monitoring of globular clusters has only detected 8 persistent and 
13 transient X-ray binaries in globular clusters to date \citep{Bahramian2014,Sanna17,Sanna18}.  Thus, we conclude that we have seen outbursts from less than 16\% of the quiescent NS LMXB population.  For the clusters which have the largest numbers of known quiescent NS LMXBs (NGC~6440, Ter~5, and NGC~6266), as well as many of the other such clusters, the distances are small enough that outbursts above $3\times10^{34}$ erg/sec would have been detected with MAXI, for these systems to be explained as VFXTs, they would need to be at the faint end of the distribution of VFXTs.

\section{Future prospects}
The discussion above indicates that observational work has found only a small fraction of the NS LMXBs in the Galaxy.  This is likely to change in the near future, due to the combination of eROSITA \citep{eROSITA} and the Legacy Survey of Space and Time (LSST) of the Vera Rubin Observatory \citep{2008SerAJ.176....1I}, which will be able to detect the thermal emission from a substantial subset of quiescent NS LMXBs, and the ellipsoidal modulations from their donor stars, respectively \citep{Maccarone2019}.  While these two projects will not produce a complete sample of all the NS LMXBs in the Galaxy, they should be very effective at detecting these systems over the whole Galactic Bulge, and much of the foreground of the Galactic disc, so that the full population can be modelled.

It is also of value to identify further tests of the nature of the Galactic Center sources, since, while we have invalidated the evidence favoring that they are black holes, we have not proved that they must be NSs.  With current instrumentation, the strongest tool in the arsenal for identifying the nature of the compact objects in these systems is radio emission.  The ratio of radio-to-X-ray flux for black hole X-ray binaries is considerably higher than for NS LMXBs \citep{2001MNRAS.324..923F}. Radio measurements can thus help break the degeneracies between the classes of X-ray binaries in the Galactic Center \citep{tjm2005}.  While some outliers to this correlation might exist, it remains the most straightforward manner in which to approach this problem at the present time, and one on which the ngVLA should be able to improve further \citep{ngvla}.\footnote{The Square Kilometre Array can do this type of work further out from the Galactic Centre\citep{CorbelSka}, but given its lack of high frequency coverage it is likely to be far more challenging for SKA than ngVLA in the innermost parts of the Galaxy.}

{ In the future, X-ray astronomy may also benefit from higher duty cycles of sensitive observations, which may open up the possibility of detecting fainter outbursts than have been seen to date with wide field instruments.  Such outbursts are likely to be qualitatively different in nature than standard outbursts due to the ionization instability, because a minimum amount of mass needs to be involved to trigger the ionization instability.  In recent years, a few accreting white dwarfs {\it have} shown long, faint outbursts in which the optically thick components of their accretion discs never reached high enough temperatures for the disc to become ionized \citep{Rivera2020,Rivera2021,Sunny}, and these are likely due to changes in the mass transfer rate from the donor stars, rather than due to disc processes \citep{Rivera2021}.  This is likely to happen in X-ray binaries as well, and such long term, low amplitude variability is already well-noted in the optical bands in one black hole X-ray binary \citep{Cantrell}, and in the X-ray band in at least one accreting neutron star \citep{Cackett} and one black hole \citep{Bernardini}.  This high amplitude variability in quiescence thus does not offer any more diagnostic power of whether the accretor is a black hole or a neutron star than the {\it bona fide} outbursts.  It {\it is}, on the other hand, reasonably likely that if high duty cycle monitoring were available of the whole sky down to the quiescent X-ray fluxes of typical LMXBs that some variability pattern could emerge that might be useful for distinguishing between black holes and neutron stars, but it is unlikely that this would be a mere reduction in the recurrence times only for the neutron star X-ray binaries.
}

{ \section{Summary}

Our primary goal in this paper is to establish that the absence of large outbursts from a quiescent X-ray binary in a 10 year timespan is not evidence for the object being a black hole.  Rather, such objects may be black holes or neutron stars, and the lack of outbursts provides no evidence about the nature of the compact accretor.

First, in Section 2, we have shown that even from the known transient neutron star X-ray binaries, at least 40\%  of the objects have recurrence times longer than 10 years, and about 1/4 of the objects have shown only a single outburst in the 60 year history of X-ray astronomy.  This point alone shows that the lack of outbursts from an object over a 5-10 year timespan cannot be used as evidence that the object must be a black hole.

Second, in Section 3, we have shown that reconciling the populations of millisecond radio pulsars with their neutron star low mass X-ray binary progenitors requires that the underlying population of neutron star low mass X-ray binaries which at least occasionally have large outbursts (as required to have enough accretion to spin up the pulsars) is at least 5 times as large as the known population of neutron star low mass X-ray binaries.  These systems thus must be undiscovered transients.  Furthermore, given the orbital period distribution mismatch between the observed neutron star X-ray binaries and the observed millisecond pulsars, a substantial fraction of the missing neutron star X-ray binaries must be systems with orbital periods of several days or more, a class of system which generically has shown single outbursts among the known transients, and which is expected to show recurrence times of millenia from theory \citep{Piro2002}.  Notably, these long period, long duration, long recurrence time outbursts are expected to be especially bright in quiescence as well, making this class' properties a good match to the observed properties of the Galactic Center quiescent X-ray binaries.

Finally, in section 4, we have shown that the Milky Way's globular cluster system has about 50 known quiescent neutron star LMXBs, and is expect to have about 130 of these objects.  Only 10\% of these have shown outbursts during the history of X-ray astronomy, indicating, again, that there must be a large underlying population of rare outbursters.

}
\section*{Data Availability}

This paper presents re-analysis of existing data which can be obtained from published papers, the Vizier service, or NASA archives for Swift and RXTE or JAXA archives for MAXI.



\bibliographystyle{mnras}
\bibliography{example} 








\bsp	
\label{lastpage}
\end{document}